\documentclass[preprintnumbers, floatfix,letterpaper,aps,prd,epsfig,nofootinbib,
twocolumn
]{revtex4-1}

\usepackage{bm,graphicx,dcolumn,epstopdf,epsf, latexsym,mathbbol, amssymb,amsmath,color,slashed, mathrsfs,mathcomp,simplewick}
\usepackage[center]{subfigure}
\usepackage{multirow}
\usepackage{makecell}
\usepackage[colorlinks,linkcolor=blue,citecolor=blue,urlcolor=blue]{hyperref}
\begin{document}
\allowdisplaybreaks
%%%%%%%%%%%%%%%%%%%%%%%%
 \newcommand{\bq}{\begin{equation}}
 \newcommand{\eq}{\end{equation}}
 \newcommand{\bqn}{\begin{eqnarray}}
 \newcommand{\eqn}{\end{eqnarray}}
 \newcommand{\nb}{\nonumber}
 \newcommand{\lb}{\label}
 \newcommand{\f}{\frac}
 \newcommand{\p}{\partial}
%%%%%%%%%%%%%%%%%%%%%%%%%
\newcommand{\PRL}{Phys. Rev. Lett.}
\newcommand{\PLB}{Phys. Lett. B}
\newcommand{\PRD}{Phys. Rev. D}
\newcommand{\CQG}{Class. Quantum Grav.}
\newcommand{\JCAP}{J. Cosmol. Astropart. Phys.}
\newcommand{\JHEP}{J. High. Energy. Phys.}
\newcommand{\red}{\textcolor{black}}
 %%%%%%%%%%%%%%%%%%%%%%%%
%

\title{Constraints on the Nieh-Yan modified teleparallel gravity with gravitational waves}

\author{Qiang Wu${}^{a, b}$}

\author{Tao Zhu${}^{a, b}$}
\email{corresponding author:  zhut05@zjut.edu.cn}

\author{Rui Niu${}^{c, d}$}

\author{Wen Zhao${}^{c, d}$}
\email{wzhao7@ustc.edu.cn}

\author{Anzhong Wang${}^{e}$}
\email{anzhong$\_$wang@baylor.edu}

\affiliation{
${}^{a}$ Institute for Theoretical Physics and Cosmology, Zhejiang University of Technology, Hangzhou, 310032, China\\
${}^{b}$ United Center for Gravitational Wave Physics (UCGWP), Zhejiang University of Technology, Hangzhou, 310032, China\\
${}^{c}$ CAS Key Laboratory for Research in Galaxies and Cosmology, Department of Astronomy, University of Science and Technology of China, Hefei 230026, China; \\
${}^{d}$ School of Astronomy and Space Sciences, University of Science and Technology of China, Hefei, 230026, China;\\
${}^{e}$ GCAP-CASPER, Physics Department, Baylor University, Waco, Texas 76798-7316, USA}

\date{\today}

\begin{abstract}

The discovery of gravitational waves (GWs) from the compact binary components by LIGO/Virgo Collaboration provides an unprecedented opportunity for testing gravity in the strong and highly dynamical field regime of gravity.  Currently a lot of model-independent tests have been performed by LIGO/Virgo Collaboration and no any significant derivation from general relativity has been found. In this paper, we study the parity violating effects on the propagation of GWs in the Nieh-Yan modified teleparallel gravity, a theory which modifies general relativity by a parity violating Nieh-Yan term. We calculate the corresponding parity violating  waveform of GWs produced by the coalescence of compact binaries. By comparing the two circular polarization modes, we find the effects of the velocity birefringence of GWs in their propagation caused by the parity violation due to the Nieh-Yan term, which are explicitly presented in the GW waveforms by the phase modification.  With such phase modifications to the waveform, we perform the full Bayesian inference with the help of the open source software {\texttt Bilby} on the GW events of binary black hole merges (BBH) in the LIGO-Virgo catalogs GWTC-1 and GWTC-2. We do not find any significant evidence of parity violation due to the parity violating Nieh-Yan term and thus place an upper bound on the energy scale $M_{\rm PV} <  6.5 \times 10^{-42} \; {\rm GeV}$ at 90\% confidence level, which represents the first constraint on the Nieh-Yan modified teleparallel gravity so far.

\end{abstract}

\maketitle

\section{Introduction}
\renewcommand{\theequation}{1.\arabic{equation}} \setcounter{equation}{0}

The direct detections of gravitational waves (GWs) emitted from the compact binary components (CBC) by LIGO/Virgo Collaboration open a new era for exploring the nature of gravity in the strong and highly-dynamical field regime of gravity \cite{gw150914, gw170817, gw-other, LIGOScientific:2017ycc}.  To data, there are 50 GW events that have been reported by  Advanced Laser Interferometer Gravitational-Wave Observatory (LIGO) and Advanced Virgo, which are included in the Gravitational-Wave Transient Catalogs GWTC-1 \cite{gwtc1} and GWTC-2 \cite{gwtc2}. In addition, two new GW events, GW200105 and GW200115, have been released recently, which most likely, are signals from mergers of neutron star and black hole binaries \cite{LIGOScientific:2021qlt}. With these events, various model-independent tests of general relativity (GR) have been performed by LIGO/Virgo Collaboration and no any significant derivation from GR has been found \cite{gw150914-testGR, gw170817-testGR, gw170817-speed, testGR_GWTC1, testGR_GWTC2}. 

Although GR has been considered to be the most successful theory of gravity since it was proposed, it faces difficulties both theoretically (e.g., singularity, quantization, etc), and observationally (e.g., dark matter, dark energy, etc).  Various modified gravities  have been proposed to be one of the effective ways to solve these anomalies \cite{MG1, MG2, MG3, MG4}. Therefore, the tests of the modified gravities are essential to confirm the final theory of gravity. 

As is well known, symmetry permeates nature and is important to all laws of physics. Thus, one important aspect for tests of gravity is to test its symmetries. Parity symmetry is one of the fundamental symmetries of GR. However, it is well known that nature is parity violating, since the first discovery of parity violation in weak interactions \cite{parity_violation}. On the other hand, when one considers the quantization of gravity, such symmetry could be also violated at high energy regimes. For examples, the parity violations in gravity can in general arise from the gravitational anomaly of the standard model of elementary particles, the Green-Schwarz anomaly canceling mechanism in string theory, or the scalarization of the Barbero-Immirzi parameter in the loop quantum gravity \cite{cs1, cs2}, see also \cite{cs_review} for a review. In this sense, the parity symmetry can only be treated as an approximate symmetry, which emerges at low energies and is violated at higher energies. With these thoughts, a lot of modified gravity theories or phenomenological models with parity violation in the gravitational interaction have been proposed, such as the Chern-Simons modified gravity \cite{cs_review, chern-simons1, chern-simons2, chern-simons3, chern-simons4, chern-simons4, chern-simons5}, the symmetric teleparallel equivalence of GR theory \cite{Conroy}, Horava-Lifshitz theories of quantum gravity \cite{horava1, horava2, horava3, horava4, horava5, horava6}, chiral scalar-tensor theory \cite{chiral_ST, chiral_ST1, chiral_ST2}, and the standard model extension \cite{SME1,SME2, SME3, SME4, SME5}. 

Parity symmetry implies that a directional flipping to the left and right does not change the laws of physics. The parity violation in gravity in general can induce an asymmetry of the propagation speed and amplitude damping between left- and right-hand polarizations of a GW, which leads to the velocity and amplitude birefringence, respectively. In primordial cosmology, such birefringence phenomenons can produce circularly polarized primordial GWs, which leaves the significant imprints in the temperature and polarization anisotropies of cosmic microwave background radiation (CMB) \cite{horava3, horava4, horava5, PGW1, PGW2, PGW3, Fu:2020tlw}. The detection of GW  emitted from the compact binary components by LIGO-Virgo provides a great opportunity to test the parity violation in gravity as well. A lot of tests on both the velocity and amplitude birefringence of GWs have been carried out by using the observational data from GW events in LIGO-Virgo catalogs \cite{SME1, chiral_ST1, CS_gb, SME4, SME5, Okounkova:2021xjv, Hu:2020rub, sai_wang, tanaka}. Recently, the parity violating effects which induces velocity birefringence due to the leading-order higher derivative modification in the waveform has been constrained through the Bayesian parameter estimation on the third Open Gravitational-wave Catalog events \cite{yi-fan1, Wang:2021gqm}. The parity violating energy scale due to the leading-order higher derivative modification has been constrained to be $M_{\rm PV} > 0.14 \; {\rm GeV}$ at 90\% confidence level, which represents the tightest bound on $M_{\rm PV}$ so far. Here the $M_{\rm PV}$ denotes the parity violating energy scale from leading high derivative corrections which corresponds to the case with $\beta_\mu=1$ as defined in \cite{waveform}. It is worth noting that the parity violating energy scale $M_{\rm PV}$ in this paper corresponds to $\beta_\mu = -1$ \cite{waveform}.

Recently, a new parity violating gravity model, the Nieh-Yan modified teleparallel gravity, was proposed in \cite{Li:2020xjt, Li:2021wij}. This model is healthy and simple in form. The Nieh-Yan modified teleparallel gravity is constructed based on the theory of GR equivalent teleparallel gravity (TEGR) \cite{TEGR} which is formulated in flat spacetime with vanishing curvature and vanishing nonmetricity (see Ref. \cite{Bahamonde:2021gfp} for a recent review). The TEGR is equivalent to GR and the  Nieh-Yan modified teleparallel gravity modified TEGR by including an extra Nieh-Yan term into the gravitational action, which breaks the parity symmetry in gravity. It is interesting to mention here that such a coupling can also appear in the mechanisms \cite{NY-1, NY-2} to regularize the infinities in theories of the Einstein-Cartan manifold, similar to the QCD axion coupling in the Peccei-Quinn mechanism \cite{NY-3} for a solution to the strong CP problem. In contrast to other parity violating gravities which break parity due to high-order derivative terms, the Nieh-Yan modified teleparallel gravity has no higher derivatives and successfully avoids the ghost mode. The cosmological perturbations and parametrized post-Newtonian limit in this theory have also been explored recently in \cite{Li:2020xjt, Li:2021wij, PPN}. Some other modified theories in the framework of teleparallel gravity or with Nieh-Yan term and their implications in the GW observations have also been considered in \cite{Bombacigno:2021bpk, Bahamonde:2017wwk, Bahamonde:2015zma, Hohmann:2020dgy, Zhang:2021kqn}.

Since no higher derivative is introduced in the Nieh-Yan modified teleparallel gravity, the effects of parity violation due to the Nieh-Yan term on the propagation of GWs is in the lower energy regime. This is in contrast to those parity violating gravities due to high derivative terms, while the propagation of GW is modified at high energy regime. This implies that the effect of Nieh-Yan term is more sensitive to the low frequency GWs. Another important property of GW in this model is that it only leads to velocity  birefringence and there is no amplitude birefringence. Similar properties of GW can also arise from the low dimension modifications to GR in the symmetric teleparallel gravity \cite{Li:2021mdp, Conroy} and the linear gravity of standard model extension \cite{SME1, SME3, SMExx} \footnote{In Refs. \cite{SME1, SME3}, only the operators with dimension $d=5$ and $d=6$ are considered. However, as examples, the Nieh-Yan term considered in this paper or lower dimension term in \cite{Li:2021mdp, Conroy} show that the inclusion of operators with lower dimension $d=3$ in the linear gravity of the SME is also possible \cite{SMExx}. It is worth noting that such lower dimension operators have been also considered in the standard model extension of electrodynamics \cite{SME-EM}.}.  In this paper, we study in detail the effects of the velocity birefringence due to the parity violating Nieh-Yan term on the GWs waveform. Decomposing the GWs into the left-hand and right-hand circular polarization modes, we find that the effects of velocity  birefringence can be explicitly presented by the modifications in the GW phase. We also mapped such phase modification to the parametrized description of parity violating waveforms proposed in \cite{waveform}.  With the modified waveform, we perform the full Bayesian inference with the help of the open source software {\bf Bilby} on the 46 GW events of BBH in the LIGO-Virgo catalogs GWTC-1 and GWTC-2. From our analysis, we do not find any signatures of parity violation due to the parity violating Nieh-Yan term and then place an upper bound on the energy scale $M_{\rm PV}$ to be $M_{\rm PV} < 6.5 \times 10^{-42}\; {\rm GeV}$ at 90\% confidence level, which represents the first constraint on the Nieh-Yan modified teleparallel gravity so far.

This paper is organized as follows. In the next section, we present a brief introduction of the Nieh-Yan modified teleparallel gravity model and then discuss the propagation of GWs in Sec.~III. In Sec. IV, we discuss the velocity birefringence effects of GWs, and then calculate the waveform of GWs produced by the coalescence of compact binary systems and particularly focus on the deviations from those in GR. In Sec.~V, we present the basic statistical framework of Bayesian analysis used in this work and report the results of constraints on the Nieh-Yan modified teleparallel gravity from the Bayesian analysis. We finish with concluding remarks and discussions in Sec.~VI.

Throughout this paper, the metric convention is chosen as $(-,+,+,+)$, greek indices $(\mu,\nu,\cdot\cdot\cdot)$ \red{and latin indices $(a, b, c, \cdot\cdot\cdot)$ which run over $0,1,2,3$ denote the spacetime and tangents space respectively, and the latin and latin indices $(i, \; j,\;k, \; l, \cdot\cdot\cdot)$ which run over $1, 2, 3$ indicate the spatial index.} We choose the units $G =c=1$.

\section{ The Nieh-Yan modified teleparallel gravity}
\renewcommand{\theequation}{2.\arabic{equation}} \setcounter{equation}{0}

In this section, we present a brief introduction of the  Nieh-Yan modified teleparallel gravity, for details about this theory, see \cite{Li:2020xjt, Li:2021wij} and references therein. 

The Nieh-Yan modified teleparallel gravity is constructed based on the theory of teleparallel gravity (TEGR) \cite{TEGR} which is equivalent to GR but formulated in flat spacetime with vanishing curvature and vanishing nonmetricity. In this theory, the dynamical variable is the tetrad field $e_a^\mu$.
%, where Latin and Greek index indicate tangents space and space-time index, respectively. 
The relation between the metric $g_{\mu\nu}$ and the tetrad field $e^a_\mu$ read
\bqn
g_{\mu\nu} = e_\mu^a e_\nu^a \eta_{ab},
\eqn
where $\eta_{ab}= (-1, 1, 1, 1)$ is the Minkowski metric. Therefore at each point of the spacetime, the tetrad field follows an orthonormal \red{basis} for the tangent field. In the construction of the TEGR, one normally starts from a flat spacetime where the curvature of the spacetime vanishes and the gravity is encoded by a nonzero torsion tensor. In this way, the torsion tensor general depends on both the tetrad field and the spin connection,
\bqn
\mathcal{T}^\lambda_{\mu\nu} = 2 e^\lambda_a (\partial_{[\mu} e_{\nu]}^a + \omega^a_{b[\mu} e^b_{\nu]}),
\eqn
where $\omega^a_{b \mu}$ is the spin connection \red{which is related to the Lorentz transformation matrix by
\bqn
\omega^a_{\; b\mu} = (\Lambda^{-1})^{a}_{\; c} \partial_\mu \Lambda^{c}_{\; b},
\eqn
and $\omega_{ab\mu}=-\omega_{ba\mu}$ with $\Lambda^a_{\; b}$ representing the element of an arbitrary Lorentz transformation matrix. Here $\Lambda^a_{\; b}$ is position dependent and satisfies the relation $\eta_{ab} \Lambda^{a}_{\; c} \Lambda^{b}_{\; d} = \eta_{cd}$.
}
With the tetrad field and the torsion tensor, the action of the TEGR can be written as
\bqn
S_{\rm TEGR} &=& \frac{1}{2}\int d^4x ~ e \mathcal{T} \nb\\
&\equiv & \int d^4x~ e \Bigg(-\frac{1}{2}\mathcal{T}_{\mu}\mathcal{T}^{\mu}+\frac{1}{8}\mathcal{T}_{\alpha\beta\mu}\mathcal{T}^{\alpha\beta\mu}\nb\\
&&~~~~ +\frac{1}{4}\mathcal{T}_{\alpha\beta\mu}\mathcal{T}^{\beta\alpha\mu}\Bigg), \lb{TEGR}
\eqn
where $e= {\rm det} (e^a_\mu) = \sqrt{-g}$ is the determinant of the tetrad, $\mathcal{T}_\mu = \mathcal{T}^\nu_{\mu\nu}$ is the torsion vector, and $\mathcal{T}$ is the torsion scalar. It is interesting to note that the torsion scalar is related to the Ricci scalar by
\bqn
R = - \mathcal{T} + \frac{2}{e} \partial_{\mu} (e \mathcal{T}^\mu). 
\eqn
Therefore, the action of the TEGR (\ref{TEGR}) is identical to the Einstein-Hilbert action up to a surface term,
\bqn
S_{\rm TEGR} = \int d^4x \sqrt{-g} \left[-\frac{1}{2} R(e) - \nabla_\mu \mathcal{T}^\mu\right]. \lb{TEGR_A}
\eqn

In the Nieh-Yan modified teleparallel gravity, the theory of TEGR is modified by introducing a Nieh-Yan term into the TEGR action \cite{Li:2021wij, Li:2020xjt}, i.e,
\bqn\label{NY1}
S_{\rm NY}=\frac{\red{c} }{4}\int d^4x \sqrt{-g}  \theta\,\mathcal{T}_{a\mu\nu}\widetilde{\mathcal{T}}^{a\mu\nu},
\eqn
where $c$ is the coupling constant, $\widetilde{\mathcal{T}}^{a\mu\nu}=(1/2)\varepsilon^{\mu\nu\rho\sigma}\mathcal{T}^a_{~~\rho\sigma}$ is the dual of the torsion two form with $\mathcal{T}^a_{\mu\nu} = 2 (\partial_{[\mu} e_{\nu]}^a + \omega^a_{b[\mu} e^b_{\nu]})$ and $\varepsilon^{\mu\nu\rho\sigma}$ being the Levi-Civita tensor. As mentioned in \cite{Li:2020xjt}, the Nieh-Yan term is itself a topological term and thus it does not contribute to the gravitational dynamics. \red{In general, the Nieh-Yan term can be split into two individual parity violating terms \cite{Hohmann:2020dgy}. To incorporate the parity violation with this two terms in the gravitational dynamics, only one of them can be added into the action. As shown in \cite{PGG_pv}, however, including either terms in the theory leads to a propagating ghost mode so that such theory is not heathy. Another} way to incorporate the parity violating effects is to consider a coupling between the Nieh-Yan term and a scalar field $\theta$. With such a scalar field, the introduction of the Nieh-Yan term (\ref{NY1}) into the action breaks the parity symmetry of the gravitational interaction. It is shown in \cite{Li:2020xjt} through perturbative analysis that such simple theory is ghost-free and healthy.  As we already mentioned in the Introduction, such a term can also appear in the mechanisms \cite{NY-1, NY-2} to regularize the infinities in theories of the Einstein-Cartan manifold, similar to the QCD axion coupling in the Peccei-Quinn mechanism \cite{NY-3} for a solution to the strong CP problem. 

By taking into account both the kinetic and potential terms of the scalar field, the full action of the Nieh-Yan modified teleparallel gravity is
\bqn\label{oldmodel}
S&=&S_{\rm TEGR} + S_{\rm NY} + S_{\theta} + S_{\rm m} \nb\\
& = &\int d^4x \sqrt{-g}\Bigg[-\frac{R(e)}{2}
+\frac{c }{4}\,\theta\,\mathcal{T}_{A\mu\nu}\widetilde{\mathcal{T}}^{A\mu\nu} \nb\\
&&~~~~~~~~ +\frac{\mathfrak{b}}{2}\nabla_{\mu}\theta\nabla^{\mu}\theta-\mathfrak{b}V(\theta)\Bigg]+S_{\rm m},
\eqn
where $\mathfrak{b}$ is a coupling constant, the curvature scalar $R(e)$ is defined by the Levi-Civita connection and considered as being fully constructed from the metric, and in turn from the tetrad. In writing the above action, we have dropped all the surface terms arising in (\ref{TEGR_A}) and (\ref{NY1}). At classical level, surface terms do not contribute to the gravitational dynamics and thus will not affect the analysis presented in this paper. It is worth mention that the nondynamical surface terms in the gravitational action can play essential roles in  the phase integral approach to quantum gravity,  the calculation of black hole entropy, and the interpretation of energy or mass in gravity \cite{Jimenez:2021nup}, etc. 

Then variation of the action with respect to the tetrad field $e^{a}_{\mu}$ and Lorentz matrix element $\Lambda^{a}_{\;\;b}$, one obtains,
\bqn
 G^{\mu\nu}+N^{\mu\nu}&=&T^{\mu\nu}+T^{\mu\nu}_{\theta}~,\label{eom1}\\
 N^{[\mu\nu]}&=&0, \label{eom2}
\eqn
where $G^{\mu\nu}$ is the Einstein tensor, $T^{\mu\nu}=-(2/\sqrt{-g})(\delta S_m/\delta g_{\mu\nu})$ and $T^{\mu\nu}_{\theta}= \mathfrak{b}[V(\theta)-\nabla_{\alpha}\theta\nabla^{\alpha}\theta/2]g^{\mu\nu}+\nabla^{\mu}\theta\nabla^{\nu}\theta$ are the energy-momentum tensors for the matter and the scalar field $\theta$ respectively, and $N^{\mu \nu}=c\, e_{a}^{~\,\nu} \partial_{\rho} \theta\, \widetilde{\mathcal{T}}^{a \mu \rho}$. Variation of the action with respect to the scalar field $\theta$ leads to the equation of motion for the scalar field, which is
\bqn
\mathfrak{b} \Big[\nabla_\mu \nabla^\mu \theta + V'(\theta) \Big]- \frac{c}{4} \mathcal{T}_{a \mu\nu} \tilde T^{a \mu\nu} =0.
\eqn
Here $V'(\theta) = dV(\theta)/d\theta$. 

\red{It is mentioned in \cite{Li:2020xjt, Li:2021wij} that Eq.~(\ref{eom2}) is the antisymmetric part of Eq.~(\ref{eom1}). From (\ref{eom2}), it is evident that the antisymmetric part of the field equations simply vanishes thus there is no antisymmetric degrees of freedom in this theory. This leads to six strong constraints on the theory. The theory contains 16+6 basic variables, 16 in the tetrad fields $e_{\mu}^a$ and 6 in the Lorentz matrix $\Lambda^a_{\; b}$. In the meantime, the theory also has 4 spacetime diffeomorphisms and 6 local Lorentz symmetries. Considering the 6 additional constraints given by (\ref{eom2}), the theory can have the same physical degrees of freedom as that in GR. At the linear perturbative level, it is shown \cite{Li:2021wij} that both the scalar and vector modes are not dynamical degrees of freedom and only two dynamical tensorial modes exist. It is worth mentioning that the degrees of freedom of the theory may also be hidden in certain backgrounds. This is also known as the strong coupling problem if such hidden modes do exist \cite{strong}. It implies that the perturbative analysis around specific background geometries cannot be fully trusted \cite{strong}. The phenomenon of degrees of freedom being hidden under special backgrounds also appears in $f(T)$ models \cite{Ong:2013qja}, the models of massive gravity \cite{DeFelice:2012mx}, and the Einsteinian cubic gravity \cite{Jimenez:2020gbw}.}

Similar to the Chern-Simons modified gravity, there are two different versions of the Nieh-Yan modified teleparallel gravity, the dynamical version and nondynamical version. The dynamical version corresponds to  $\mathfrak{b} \neq 0$, while the non-dynamical version corresponds to $\mathfrak{b}=0$. When $\mathfrak{b} =0$, the equations of motion now reduce to
\bqn
G^{\mu\nu}+N^{\mu\nu}&=&T^{\mu\nu}, \label{eom3}\\
 N^{[\mu\nu]}&=&0~,\label{eom4}
\eqn
and the equation of motion for the scalar field now becomes a constraint
\bqn
 \mathcal{T}_{a \mu\nu} \tilde T^{a \mu\nu}=0.
\eqn
As shown in \cite{Li:2020xjt, Li:2021wij}, the propagation of GWs in both theories follows the same propagating equation, thus in this paper we will not distinguish this two versions and take $\mathfrak{b}=1$ hereafter for simplification.

\section{GWs in the Nieh-Yan modified teleparallel gravity}
\renewcommand{\theequation}{3.\arabic{equation}} \setcounter{equation}{0}

Let us investigate the propagation of GW in the Nieh-Yan modified teleparallel gravity with the action given by (\ref{oldmodel}). \red{According to Eq.~(\ref{eom2}), the theory does not contain the antisymmetric perturbations thus we only focus on the symmetric tensorial perturbations, i.e., two independent modes of GW.} We consider the GWs propagating on a homogeneous and isotropic background. The spatial metric in the flat Friedmann- Robertson-Walker universe is written as
\bqn
g_{ij} = a(\tau) (\delta_{ij} + h_{ij}(\tau, x^i)), \lb{metric_spatia}
\eqn
where $\tau$ denotes the conformal time, which relates to the cosmic time $t$ by $dt =a d\tau$, and $a$ is the scale factor of the universe. Throughout this paper, we set the present scale factor $a_0 =1$. $h_{ij}$ denotes the GWs, which represents the transverse and traceless metric perturbations, i.e., 
\bqn
\partial^i h_{ij} =0 = h^i_i.
\eqn
The above spatial metric in (\ref{metric_spatia}) corresponds to the perturbations of the tetrad fields as
\bqn
e_0^0 = a, e_{i}^0 = 0, e^{a}_0 =0, \\
e_{i}^a = a \left(  \gamma_i^a  + \frac{1}{2}\gamma^{aj} h_{ij} \right),
 \eqn
 where $\gamma_i^a$ can be regarded as the spatial tetrad on three-dimensional spatial hypersurface. For a flat universe one has $\delta_{ij} = \delta_{ab} \gamma^a_i \gamma^b_j$. Here we would like to mention that the tensor perturbations only come from the tetrad field, and the spin connection or the local Lorentz matrices do not contribute to the tensor perturbations. 
 
 To proceed further one can substitute the above tetrad fields in to the action (\ref{oldmodel}) and expand  the second order in $h_{ij}$. After tedious calculations, one finds \cite{Li:2020xjt, Li:2021wij},
 \bqn
 S^{(2)} = \int d^4 x \frac{a^2}{8} \left(h'_{ij} h'_{ij}  - \partial_k h_{ij} \partial^k h^{ij} - c \theta' \epsilon_{ijk} h_{il} \partial_j h_{kl}\right),\nb\\
 \eqn
 where $\epsilon_{ijk}$ is the antisymmetric symbol and a prime denotes the derivative with respect to the conformal time $\tau$. We consider the GWs propagating in the vacuum, and ignore the source term. Varying the action with respect to $h_{ij}$, we obtain
 \bqn
 h''_{ij} + 2 \mathcal{H} h'_{ij} - \partial^2 h_{ij} + \frac{1}{2}c \theta' ( \epsilon_{lki} \partial_l h_{jk}  +  \epsilon_{lkj} \partial_l h_{ik})=0,\nb\\
 \eqn
 where $\mathcal{H} \equiv a'/a$.
 
 In the parity-violating gravities, it is convenient to decompose the GWs into the circular polarization modes. To study the evolution of $h_{ij}$, we expand it over spatial Fourier harmonics,
 \bqn
 h_{ij}(\tau, x^i) = \sum_{A={\rm R, L}} \int \frac{d^3k}{(2\pi)^3} h_A(\tau, k^i) e^{i k_i x^i} e_{ij}^A(k^i),\nb\\
 \eqn
 where $e_{ij}^A$ denote the circular polarization tensors and satisfy the relation
 \bqn
 \epsilon^{ijk} n_i e_{kl}^A = i \rho_A e^{j A}_l,
 \eqn
 with $\rho_{\rm R} =1$ and $\rho_{\rm L} = -1$. We find that the propagation equations of these two modes are decoupled, which can be casted into the standard parametrized form proposed in \cite{waveform},
 \bqn\lb{eom_A}
 h''_A + (2+\nu_A) \mathcal{H} h'_A + (1+ \mu_A) k^2 h_A=0,
 \eqn
 where
 \bqn
 \nu_A =0,  \;\;\;  \mu_A = \frac{ \rho_A c \theta' }{k}. 
 \eqn
 In the above parametrization, the effects of the parity violation are fully characterized by two parameters: $\nu_A$ and $\mu_A$. As shown in refs. \cite{waveform}, such parametrization provides an unifying description for the low-energy effective description of GWs in generic parity violating gravities, including Chern-Simons modified gravity, ghost-free parity violating scalar-tensor theory, symmetric teleparallel equivalence of general relativity, Ho\v{r}ava-Lifshitz gravities, and the Nieh-Yan modified  teleparallel gravity.  The parameter $\mu_A$ leads to different velocities of left-hand and right- hand circular polarizations of GWs, so that the arrival times of the two circular polarization modes could be different. The parameter $\nu_A$, on the other hand, leads to different damping rates of  left-hand and right- hand circular polarizations of GWs, so that the amplitude of left-hand circular polarization of gravitational waves will increase (or decrease) during the propagation, while the amplitude for the right-hand modes will decrease (or increase). In the Nieh-Yan modified teleparallel gravity, we have $\nu_A=0$ and $\mu_A = \rho_A c \theta' /k$, therefore there is no modification on the damping rate of GWs and the parity violation due to the Nieh-Yan term can only affect the velocities of GWs. This is the phenomenon of velocity birefringence. It is worth noting here that similar corrections on $\mu_A  \propto 1/k$ can also arise from the lower dimension operators in the parity violating symmetric teleparallel equivalence of GR theory \cite{Conroy, Li:2021mdp}.

 \section{Velocity birefringence and phase modifications to the waveform of GWs}
 \renewcommand{\theequation}{4.\arabic{equation}} \setcounter{equation}{0}
 
 In this section, we study the velocity birefringence effects during the propagation of GWs. As shown in \cite{waveform}, for each circular polarization mode $h_A$, the velocity birefringence effect induces the phase corrections to the waveform of GWs. In order to derive the phase modification, similar to our previous works \cite{waveform}, let us first define $u_A(\tau) = \frac{1}{2} a(\tau) M_{\rm Pl} h_A(\tau)$, and then the equation of motion (\ref{eom_A}) can be casted into the form
 \bqn
 \frac{d^2 u_A}{d\tau^2} + \left(\omega_A^2 - \frac{a''}{a}\right) h_A =0,
 \eqn
 where
 \bqn
 \omega_A = k^2 \left(1+  \rho_A \frac{c \theta'}{k} \right),
 \eqn
 is the modified dispersion relation.  Then, one can find that GWs with different helicities will have different phase velocities
 \bqn
 v_A \simeq 1 + \rho_A  \frac{c \theta'}{2 k}.
 \eqn
 Since $\rho_A$ has the opposite signs for left-hand and right-hand polarization modes, it is straightforward to see that the phase velocities for these two modes are different. For later convenience, one can introduce a characteristic energy scale $M_{\rm PV} = c \theta ' /a = c \dot \theta$, and then one has
 \bqn
v_A = 1 +  \rho_A \frac{a M_{\rm PV}}{2k}.
\eqn

Now considering a graviton emitted radially at $r=r_e$ and received at $r=0$, we have,
\bqn
\frac{dr}{dt} = - \frac{1}{a} \left[1 +\rho_{A}  \frac{a M_{\rm PV} }{2k} \right].
\eqn
Note that in the above we have assumed $c \dot \theta$ to be a constant. Integrating this equation from the emission time ($r=r_e$) to arrival time ($r=0$), one obtains
\bqn
r_e=  \int_{t_e}^{t_0} \frac{dt}{a(t)} +  \rho_{A}  \frac{M_{\rm PV} }{2 k } \int_{t_e}^{t_0} dt.
%}
\eqn
Consider gravitons emitted at two different times $t_e$ and $t_e'$, with wave numbers $k$ and $k'$, and received at corresponding arrival times $t_0$ and $t_0'$ ($r_e$ is  the same for both).  Assuming $\Delta t_e \equiv t_e- t_e ' \ll a/\dot a$, then the difference of their arrival times is given by 
\bqn
\Delta t_0= (1+z) \Delta t_e +\frac{ \rho_A }{2} \left( \frac{M_{\rm PV}}{k'} - \frac{M_{\rm PV}}{k}\right)\int_{t_e}^{t_0} dt, \nb\\
\lb{time}
\eqn
where $z = 1/a(t_e) -1$ is the cosmological redshift. 

Let us turn to consider the GW signal emitted from the nonspinning, quasicircular inspiral of compact binary system in the post-Newtonian approximation.  Relative to the GW in GR, the parity violation due to the Nieh-Yan term modifies the phase of GWs. In the Fourier domain, $h_A$ can be calculated analytically in the stationary phase approximation, which is given by
\bqn
h_A(f) = \frac{{\cal A}_A(f)}{\sqrt{df/dt}} e^{i \Psi(f)},
\eqn
where $f$ is the frequency at the detector and $\Psi$ denotes the phase of the GWs.  As shown in \cite{waveform}, the difference of arrival times induces the modification of GW phase $\Psi$ as follows,
\bqn
\Psi_A(f)= \Psi^{\rm GR}_A (f) + \rho_A  \delta \Psi_1(f),
\eqn
where
\bqn
\delta \Psi_1(f) = A_{\mu} \ln u \lb{deltaPsi}
\eqn
with 
\bqn
A_\mu = \frac{M_{\rm PV}}{2 H_0} \int_0^z \frac{dz}{(1+z) \sqrt{(1+z)^3 \Omega_m + \Omega_\Lambda}}. \lb{Amu}
\eqn
Here $u= \pi {\cal M} f$ with $f = k/2\pi$ being the frequency of GWs and ${\cal M} = (1+z) {\cal M}_{\rm c}$, where ${\cal M}_c \equiv (m_1 m_2)^{3/5}/(m_1+m_2)^{1/5}$ is the chirp mass of the binary system with component masses $m_1$ and $m_2$. We adopt a Planck cosmology with $\Omega_m = 0.315$, $\Omega_\Lambda = 0.685$, and $H_0=67.4\; {\rm km} \; {\rm s}^{-1} \; {\rm Mpc}^{-1}$. With the above phase correction, one can write the waveform of GWs with the effects of the Nieh-Yan term in the form
\bqn
h_A(f) = h_A^{\rm GR}(f) e^{i \rho_A \delta \Psi_1}.
\eqn

In order to test the Nieh-Yan modified teleparallel gravity with observations of GWs, it is convenient to analyze the GWs in the Fourier domain. The responses of detectors for the GW signals $h (f)$ in the Fourier domain can be written in terms of waveforms of $h _{+}$ and $h_{\times}$ as
\bqn 
h (f) = [F_{+} h _{+}(f) + F_\times  h _{\times}(f)] e^{-2 i \pi f \triangle t},
\eqn 
where $F_{+}$ and $F_{\times}$ are the beam pattern functions of GW detectors, depending on the source location and polarization
angle \cite{GR_wave}. $\triangle t$ is the arrival time difference between the detector and the geocenter. In GR, the waveform of the two polarizations $h_{+}(f) $ and $h _{\times}(f) $ are given respectively by \cite{GR_wave2}
\bqn
h ^{\rm GR} _{+}
= (1 + \chi^2) \mathcal{A} e^{i \Psi},~~ 
h ^{\rm GR} _{\times} = 2 \chi \mathcal{A} e^{i( \Psi + \pi/2)},
\eqn 
where $\mathcal{A}$ and $\Psi$ denote the amplitude and phase of the waveforms $h^{\rm GR}_{+, \times}$, and $\chi = \cos \iota $ with $\iota$ being the inclination angle of the binary system. In GR, the explicit forms of $\mathcal{A}$ and $\Psi$ have been calculated in the high-order PN approximation (see for instance \cite{GR_wave2} and references therein). 

Now using the relationship between $h_{+, \times}$ and $h_{\rm R, L}$, i.e., 
\bqn
h_{+} = \frac{h_{\rm L} + h_{\rm R}}{\sqrt{2}}, \;\; h_{\times} =  \frac{h_{\rm L} - h_{\rm R}}{\sqrt{2} i},
\eqn
one can obtain the effects of the Nieh-Yan term on the above waveform for the plus and cross modes. It is not difficult to get
\bqn
h_+ = h_{+}^{\rm GR} \cos(\delta \Psi_1)+ h_{\times}^{\rm GR} \sin(\delta \Psi_1), \lb{h_plus}\\
h_\times = h_\times^{\rm GR} \cos(\delta \Psi_1)- h_{+} \sin(\delta \Psi_1).\lb{h_cross}
\eqn
Therefore, the Fourier waveform $h(f)$ becomes
\bqn
h (f) = \mathcal{A} \delta \mathcal{A} e ^{i (\Psi + \delta \Psi)},
\eqn
where
\bqn
\delta \mathcal{A} &=& \sqrt{(1 + \chi ^2)^2 F^2_{+} + 4 \chi^2 F^{2}_{\times}}  \nb\\
&& \times \left[ 1
- \frac{  (1 - \chi^2)^2 F_{+}  F_{\times}}{(1 + \chi ^2)^2 F^2_{+} + 4 \chi^2 F^{2}_{\times}} \delta \Psi_{1} \right], \nb\\
\delta \Psi &=& \tan^{-1} \left[ \frac{2 \chi F_{\times}}{(1+ \chi^2) F_{+}} \right]  \nb\\
&& +  \frac{ 2\chi (1 + \chi^2)(F^2_{+} + F^2_{\times})}{(1 + \chi ^2)^2 F^2_{+} + 4 \chi^2 F^{2}_{\times}} \delta \Psi_{1}.
\eqn

\section {Constraints on Nieh-Yan term}
 \renewcommand{\theequation}{5.\arabic{equation}} \setcounter{equation}{0}
 
 \subsection{Bayesian inference for GW data}
 
 In this paper we use the Bayesian inference to obtain the constraints on the Nieh-Yan term with selected GW events from the LIGO-Virgo catalogs GWTC-1 and GWTC-2. The Bayesian inference framework has been broadly employing in the inference of scientific conclusions  from GW data. Given a set of compact binary merger observations from GW detectors with the data $d$ of GW signals, one in general can infer the distribution of parameters $\vec{\theta}$ by comparing the predicted GW strain in each instrument with the data. One can write the Bayes' theorem in the context of GW astronomy as:
 \bqn
 P({\vec \theta}|d, H)=\frac{P(d| \vec{\theta}, H) P(\vec{\theta}| H)}{P(d|H)},\lb{bayes}
 \eqn
 where $H$ is the waveform model, $P(\vec{\theta}| H)$ is the prior distribution for model parameters $\vec{\theta}$ and $P(d| \vec{\theta}, H) $ is the likelihood for obtaining the data given a specific set of model parameters,  $P(d|H)$ is the marginalized likelihood or evidence for the model or ``hypothesis" $H$, and $ P({\vec \theta}|d, H)$ denotes the posterior probability distributions for physical parameters ${\vec{\theta}}$ which describe the observational data.

 In general, the GW signals from the compact binary mergers are extremely weak and the matched filtering method has been used to extract these signals from the noises. Considering Gaussian and stationary noise from GW detectors, the matched filtering method allows to define the likelihood function in the form of
 \bqn
 \ln P(d|{\vec \theta}, H) = -\frac{1}{2} \sum_{j=1}^N \langle d_j-h({\vec \theta})|d_j-h({\vec \theta})\rangle,
 \eqn
 where $h(\vec{\theta})$ is the \red{GW waveform template response function} in model $H$ and $j$ represents the $j$th GW detector. The noise weighted inner product $\langle A|B \rangle$ is defined as
 \bqn
 \langle A|B \rangle = 4\; {\rm Re} \left[\int_0^\infty \frac{A(f) B(f)^*}{S(f)} df\right],
 \eqn
 where $\;^*$ denotes complex conjugation and $S(f)$ is the power spectral density (PSD) function of the detector. In Bayes' theorem, the Bayes' evidence $P(d|H)$ in (\ref{bayes}) can be computed by integrating the likelihood $P(d|{\vec \theta}, H)$ of parameters ${\vec \theta}$ times the prior probabilities $P({\vec \theta}|H)$ for these parameters within the waveform model $H$,
 \bqn
 P(d|H) = \int d{\vec \theta} P(d|{\vec \theta}, H) P({\vec \theta}|H).
 \eqn

In this work, in order to constrain the Nieh-Yan modified teleparallel gravity, we employ the open-source package {\bf Bilby} to perform the Bayesian inference by analyzing the GW data from selected events of binary black hole mergers in the LIGO-Vrigo catalogs GWTC-1 and GWTC-2. For the GR waveform $h^{\rm GR}_{+, \; \times}(f)$, we use the  spin precessing waveform \texttt{IMRPhenomPv2} \cite{Schmidt:2014iyl, Hannam:2013oca} for all the BBH events  except GW190521 with the parameter vector ${\vec \theta} = \{\alpha, \delta, \psi, \phi, t_{\rm c}, d_{L}, {\cal M}, \eta, a_1, a_2, \cos\theta_1, \cos\theta_2, \phi_{12}, \phi_{\rm JL}, \theta_{JN}\}$, where $\alpha$ and $\delta$ are the right ascension and declination of the binary system in the sky, $\psi$ is the polarization angle of the source defined with respect to the Earth centered coordinates, $\phi$ is the binary phase at a reference frequency, $t_{\rm c}$ is the time of coalescence, $d_L$ is the luminosity distance to the source, ${\cal M}$ is the detector-frame chirp mass of the binary, $\eta$ is the symmetric mass ratio, $a_1 (a_2)$ is the dimensionless spin magnitude of the larger (smaller) black hole, $\theta_1 (\theta_2)$ is the angle between the spin direction and the orbital angular momentum of the binary for the larger (smaller) BH, $\phi_{12} $ is the difference between total and orbital angular momentum azimuthal angles, $\phi_{\rm JL}$ denotes the difference between the azimuthal angles of the individual spin direction projections onto the orbital plane, and $\theta_{JN}$ represents the angle between the total angular momentum and the line of sight. For GW190521, we use the state-of-the-art approximant \texttt{IMRPhenomXPHM} which includes the subdominant harmonic modes of GW and accounts for spin-precession effects for a quasicircular-orbit binary black hole coalescence \cite{Pratten:2020ceb}.

In order to constrain the parity violating Nieh-Yan term, we construct the parity-violating waveform based on the above template through (\ref{h_plus}) and (\ref{h_cross}) with $\delta \Psi_1$ being given by (\ref{deltaPsi}).  For this purpose we append one additional parameter $A_\mu$, which represents the effects of the parity violation due to the Nieh-Yan modified teleparallel gravity, in addition to the parameter vector ${\vec \theta}$. Then we consider a series of GW events comprised of data $\{d_i\}$, described by parameters $\{{\vec \theta_i}\}$, where $i$ runs from 1 to $N$ with $N$ being the number of the analyzed GW events in the Bayesian inference. Then for each event, posterior for all the parameters describing the event can be written as
\bqn
P(A_\mu, {\vec \theta_i} | d_i) = \frac{P(M_{\rm PV}, \vec{\theta}_i) P(d_i| M_{\rm PV}, \vec{\theta}_i)}{P(d_i)}.
\eqn
To infer the posterior of the parameter $A_\mu$, one can marginalize over all parameters $\vec \theta_i$ for the individual GW events. This procedure gives the marginal posterior distribution on $A_\mu$ for the $i$th GW event,
\bqn
P(A_\mu|d_i) &=& \frac{P(A_\mu)}{P(d_i)} \int P(\vec{\theta}_i) P((d_i|A_\mu, \vec{\theta}_i) d\vec{\theta_i}.
\eqn

\subsection{Results of constraints}

To data, a total of 50 GW events have been reported by the LIGO-Virgo catalogs GWTC-1 and GWTC-2. Among these events, not all of them are of our interest for the Bayesian analysis. For our purpose, we consider all 46 GW events of BBH as presented in Table.~\ref{table}.  We exclude the GW events of binary neutron star or possible binary neutron star-black hole merges, since these events are not expected to improve the constraint on $M_{\rm PV}$ drastically. The data of these 46 GW events are downloaded from the Gravitational Wave Open Science Center \cite{data_GW}. Besides strain data, power spectral densities (PSDs) are also needed for parameter estimation. Instead of directly estimating PSDs from strain data by the Welch method, we use the event-specific PSDs which are encapsulated in LVC posterior sample releases for specific events \cite{data_GW2, data_GW1}. These PSDs are expected to lead to more stable and reliable parameter estimation \cite{PSD1, PSD2}.

\begin{table}
\caption{90\% credible level upper bounds on $M_{\rm PV}$ from the Bayesian inference by  analyzing 46 GW events of BBH in the LIGO-Virgo catalogs GWTC-1 and GWTC-2.}
\lb{table}
\begin{ruledtabular}
\begin{tabular} {c|cc}
catalogs & GW events &  Constraints [$10^{-41}\; {\rm GeV}$]\\
\hline 
\multirow{10}*{GWTC-1} 
& GW150914 &  6.5 \\
& GW151012 &  6.9 \\
& GW151226 &  23.5 \\
& GW170104 & 16.2 \\
& GW170608 & 17.2 \\
& GW170729 &  2.9 \\
& GW170809 & 6.4 \\
& GW170814 & 7.2 \\
& GW170818 & 4.5 \\
& GW170823 &  3.4 \\
\hline
\multirow{36}*{GWTC-2} 
& GW190408\_181802&  3.6 \\
& GW190412 &  7.0 \\
& GW190413\_052954&  6.6 \\
& GW190413\_134308&  3.2 \\
& GW190421\_213856 &  6.9 \\
& GW190424\_180648 & 2.5 \\
& GW190503\_185404 & 3.7 \\
& GW190512\_180714 & 4.2 \\
&GW190513\_205428 & 3.8 \\
& GW190514\_065416&  3.9\\
& GW190517\_055101 & 9.1 \\
& GW190519\_153544& 5.0 \\
 & GW190521 & 4.5 \\
& GW190521\_074359 & 3.9 \\
& GW190527\_092055 & 5.1 \\
& GW190602\_175927& 3.1  \\
& GW190620\_030421&  3.4 \\
& GW190630\_185205 & 6.3 \\
& GW190701\_203306&  8.7 \\
& GW190706\_222641&   5.7\\
& GW190707\_093326&   13.2\\
& GW190708\_232457&   11.5\\
& GW190719\_215514&  7.8 \\
& GW190720\_000836&  4.4 \\
& GW190727\_060333&  3.3  \\
& GW190728\_064510&  9.1  \\
& GW190731\_140936& 4.4  \\
& GW190803\_022701&  4.4 \\
& GW190828\_063405 & 3.0 \\
& GW190828\_065509 & 9.6 \\
& GW190909\_114149&  10.3 \\
& GW190910\_112807 & 2.6 \\
& GW190915\_235702 & 3.0 \\
& GW190924\_021846 & 36.7 \\
& GW190929\_012149&  11.4 \\
& GW190930\_133541&  21.6 \\
\hline
 & Combined  &   0.65
\end{tabular}
\end{ruledtabular}
\end{table}

For these events, we perform parameter estimations using Bayesian analysis by selecting $4 \;{\rm s}$ or $8 \; {\rm s}$ data over all GW parameters $\vec{\theta}$ and the parity violating parameter $A_\mu$. The prior for the standard GW parameters ${\vec{\theta}}$ are consistent with those used in \cite{gwtc1, gwtc2}. The prior for $A_\mu$  is chosen to be uniformly distributed. We use the package {\texttt{BILBY}}  \cite{bilby} to perform the analysis and the posterior distribution is sampled by the nest sampling method dynesty over the fiducial BBH and the parity violating parameter $A_\mu$. We report our main results in the next subsection.

\begin{figure*}
{
\includegraphics[width=17.1cm]{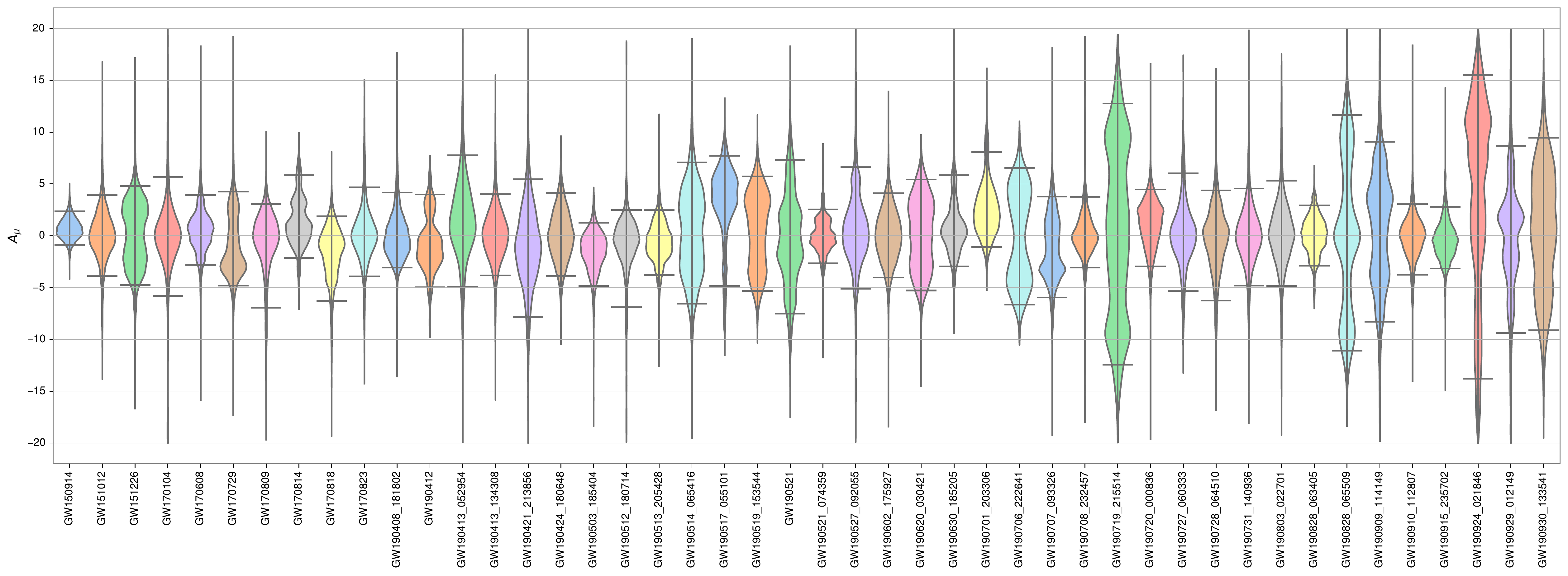}}\\
\caption{Violin plots of the posteriors of the parameter $A_\mu$. The results are obtained by analyzing the 46 GW events of BBH in the catalogs GWTC-1 and GWTC-2. The region in the posterior between the upper and lower bar denotes the $90\%$ credible interval.} \label{violin}
\end{figure*}

For all the 46 GW events we analyzed, we find that the posterior distribution of $A_{\mu}$ are all consistent with its GR value $A_\mu=0$, which means we do not find any signatures of parity violation of  Nieh-Yan modified teleparallel gravity in the data of these events. To illustrate the results of $A_\mu$ from each individual GW event, we plot Fig.~\ref{violin} to show the marginalized posterior distribution of $A_\mu$. In this figure, the region in the posterior between the upper and lower bar denotes the $90\%$ credible interval, and the bar at the middle denotes the median value. It is shown that the GR value $A_\mu=0$ is well within the 90\% confidence level for each GW event. 

\begin{figure}
{
\includegraphics[width=8.1cm]{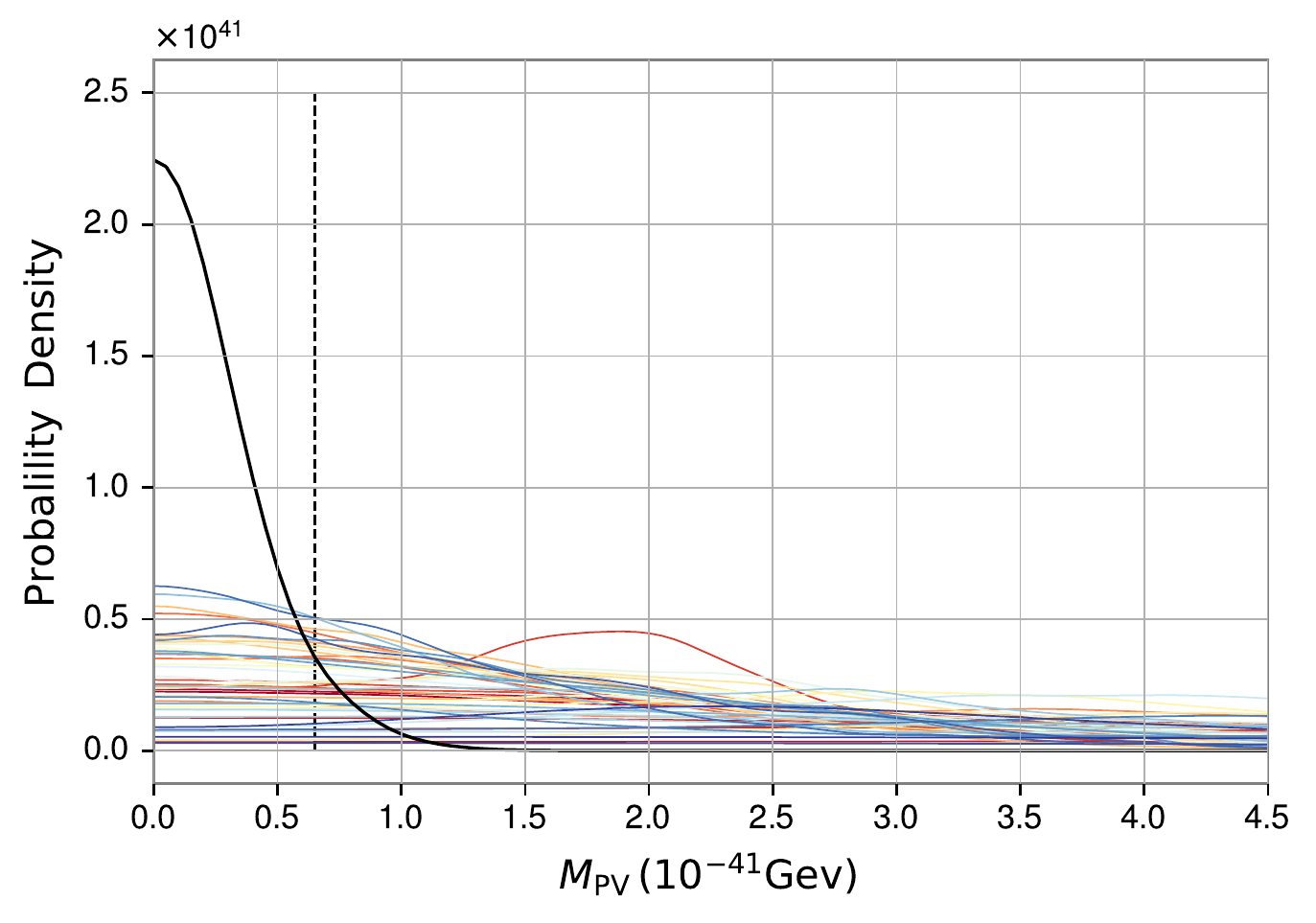}}\\
\caption{The posterior distributions for $M_{\rm PV}$ from 46 GW events in the LIGO-Virgo catalogs GWTC-1 and GWTC-2.  The solid black curve represents the combined posterior distribution of $M_{\rm PV}$ and the vertical dash line denotes the 90\% upper limits for $M_{\rm PV}$ from combined results.} \label{pdf}
\end{figure}

From the posterior distributions of $A_{\mu}$ and redshift $z$ calculated from the 46 events, one can convert $A_\mu$ and $z$ into $M_{\rm PV}$ through Eq.~(\ref{Amu}). In Fig.~\ref{pdf} we plot the marginalized posterior distribution of $M_{\rm PV}$. Then the upper bounds on $M_{\rm PV}$ for each individual event can be calculated from the corresponding posterior distribution of $M_{\rm PV}$. In Table.~\ref{table}, we present the 90\% credible level upper bounds on $M_{\rm PV}$ from the Bayesian inference of each event. From Table.~\ref{table}, one can see that the best constraints on $M_{\rm PV}$ are all from those events with ${\cal M}_{\rm c} \gtrsim 30 M_{\odot}$.

The parameter $M_{\rm PV}$ is a universal quantity for all GW events. One can combine all the individual posterior of $M_{\rm PV}$ for each event to get the overall constraint. This can be done by multiplying the posterior distributions of all these events together through 
\bqn
P(M_{\rm PV}|\{d_i\}, H) \propto \prod_{i=1}^{N} P(M_{\rm PV}| d_i, H),
\eqn
where $d_i$ denotes data of the $i$th GW event. We find that $M_{\rm PV}$ can be constrained to be
\bqn
M_{\rm PV} < 6.5 \times 10^{-42} \; {\rm GeV}
\eqn
at 90\% confidence level. We thus conclude that we do not find any significant evidence of parity violation due to the Nieh-Yan modified teleparallel gravity at $M_{\rm PV} > 6.5\times 10^{-42} \; {\rm GeV}$. This constraint in turn can convert into a constraint on parameter $c \dot \theta$ as
\bqn
c \dot \theta < 6.5 \times 10^{-42}  \; {\rm GeV}.
\eqn
So far, this upper bound represents the only observational constraint on the Nieh-Yan modified teleparallel gravity.

\section{conclusion and outlook}
 \renewcommand{\theequation}{6.\arabic{equation}} \setcounter{equation}{0}

 With the discovery of GWs from the coalescence of compact binary systems by LIGO/Virgo Collaboration, the testing of gravity in the strong gravitational fields becomes possible. Therefore, the studies of GWs in the alternative theories of gravity and inference of their constraints from data of GW events detected by LIGO/Virgo Collaboration are of crucial importance for understanding gravity under extreme conditions. In this paper, we focus on a new parity-violating gravity model, the Nieh-Yan modified teleparallel gravity \cite{Li:2020xjt, Li:2021wij}. This model is healthy and simple in form which modifies the TEGR by a parity violating Nieh-Yan term. Such a term can arise from the mechanisms to regularize the infinities in theories of the Einstein-Cartan manifold as mentioned in \cite{NY-1, NY-2}. In contrast to other parity violating gravities which break parity due to high-order derivative terms, the Nieh-Yan modified teleparallel gravity has no higher derivatives and successfully avoids the ghost mode. 
 
 In order to test this model with GWs, we study in detail the effects of the velocity birefringence due to the parity violating Nieh-Yan term on the GW waveforms. Decomposing the GWs into the left-hand and right-hand circular polarization modes, we find that the effects of velocity birefringence can be explicitly presented by the modifications in the GW phase. We also mapped such phase modification to the parametrized description of parity violating waveforms proposed in \cite{waveform}. With the modified waveform, we perform the full Bayesian inference with the help of the open source software \texttt{BILBY} on the 46 GW events of BBH in the LIGO-Virgo catalogs GWTC-1 and GWTC-2. From our analysis, we do not find any signatures of parity violation due to the parity violating Nieh-Yan term and then place an upper bound on the energy scale $M_{\rm PV}$ to be $M_{\rm PV} < 6.5 \times 10^{-42}\;  {\rm GeV}$ at 90\% confidence level, which represents the first constraints on the Nieh-Yan modified teleparallel gravity so far. 
 
 The above constraint on $M_{\rm PV}$ can be straightforwardly mapped to bound on the lower dimensional parity violating terms in the symmetric teleparallel gravity or possible operators with dimension $d=3$ in the linear gravity of the standard model extension. In the framework of the symmetric teleparallel gravity, one can modify the GR equivalent symmetric teleparallel gravity by a parity violating interaction, $ c \phi Q \tilde Q$, here $Q \tilde Q = \epsilon^{\mu\nu\rho \sigma} Q_{\mu\nu \alpha} Q_{\rho \sigma}^{\alpha}$ with $Q_{\mu\nu\sigma}$ being the nonmetricity tensor \cite{Li:2021mdp, Conroy}. It is worth noting that such a parity violating interaction can lead to the ghost problem in the vector perturbations \cite{Li:2021mdp}. The propagating equation of GWs in the  symmetric teleparallel gravity with the parity violating term $c \phi Q \tilde Q$ shares exactly the same form of (\ref{eom_A}) with $\nu_A=0$ and $\mu_A = \rho_A c  \dot \phi /(a k)$ \cite{Li:2021mdp}. This corresponds to $M_{\rm PV} = c \dot \phi$, thus one obtains
 \bqn
 |c \dot \phi | < 6.5 \times 10^{-42} \; {\rm GeV}.
 \eqn 
 
With the constant upgrading of the advanced LIGO and Virgo detectors, we expect more GW events, especially those with heavier chirp mass and higher redshifts, will be detected in the future. With more data,  we are expected to improve significantly the constraint on $M_{\rm PV}$ and the corresponding modified gravities in the future.

\section*{Acknowledgments}

T.Z., Q.W., and A.W. are supported in part by the National Key Research and Development Program of China Grant No. 2020YFC2201503, and the Zhejiang Provincial Natural Science Foundation of China under Grants No. LR21A050001 and No. LY20A050002, the National Natural Science Foundation of China under Grant No. 11675143 and No. 11975203, and the Fundamental Research Funds for the Provincial Universities of Zhejiang in China under Grant No. RF-A2019015. W.Z. and R.N. are supported by NSFC Grants No. 11773028, No. 11633001, No. 11653002, No. 11421303, No. 11903030, the Fundamental Research Funds for the Central Universities, and the Strategic Priority Research Program of the Chinese Academy of Sciences Grant No. XDB23010200.  

\appendix


\begin{thebibliography}{399}


\bibitem{gw150914}
B. P. Abbott {\em et al.} (LIGO Scientific and Virgo Collaborations), 
Observation of Gravitational Waves from a Binary Black Hole Merger,
Phys. Rev. Lett. \textbf{116}, 061102 (2016); 
GW150914: First results from the search for binary black hole coalescence with Advanced LIGO,
Phys. Rev. D \textbf{93}, 122003 (2016); 
 Properties of the Binary Black Hole Merger GW150914,
Phys. Rev. Lett. \textbf{116}, 241102 (2016); 
 GW150914: The Advanced LIGO Detectors in the Era of First Discoveries,
Phys. Rev. Lett. \textbf{116}, 131103 (2016).

\bibitem{gw170817}
B.~P.~Abbott {\em et al.} (LIGO Scientific and Virgo Collaborations),
GW170817: Observation of Gravitational Waves from a Binary Neutron Star Inspiral,
Phys. Rev. Lett. \textbf{119}, 161101 (2017).
%[arXiv:1710.05832 [gr-qc]].


\bibitem{gw-other}
B. P. Abbott {\em et al.} (LIGO Scientific and Virgo Collaborations), 
GW151226: Observation of Gravitational Waves from a 22-Solar-Mass Binary Black Hole Coalescence,
Phys. Rev. Lett. \textbf{116}, 241103 (2016); 
GW170104: Observation of a 50-Solar-Mass Binary Black Hole Coalescence at Redshift 0.2,
Phys. Rev. Lett. \textbf{118}, 221101 (2017); 
GW170608: Observation of a 19-solar-mass binary black hole coalescence,
Astrophys. J. Lett. \textbf{851}, L35 (2017).

%\cite{LIGOScientific:2017ycc}
\bibitem{LIGOScientific:2017ycc}
B.~P.~Abbott \textit{et al.} (LIGO Scientific and Virgo Collaborations), GW170814: A Three-Detector Observation of Gravitational Waves from a Binary Black Hole Coalescence,
Phys. Rev. Lett. \textbf{119}, 141101 (2017).
% doi:10.1103/PhysRevLett.119.141101
%[arXiv:1709.09660 [gr-qc]].

\bibitem{gwtc1}
B.~P.~Abbott \textit{et al.} (LIGO Scientific and Virgo Collaborations),
GWTC-1: A Gravitational-Wave Transient Catalog of Compact Binary Mergers Observed by LIGO and Virgo during the First and Second Observing Runs,
Phys. Rev. X \textbf{9}, 031040 (2019).
%doi:10.1103/PhysRevX.9.031040
%[arXiv:1811.12907 [astro-ph.HE]].

\bibitem{gwtc2}
R.~Abbott \textit{et al.} (LIGO Scientific and Virgo Collaborations),
GWTC-2: Compact Binary Coalescences Observed by LIGO and Virgo During the First Half of the Third Observing Run,
Phys. Rev. X \textbf{11}, 021053 (2021).
%doi:10.1103/PhysRevX.11.021053
%[arXiv:2010.14527 [gr-qc]].

\bibitem{LIGOScientific:2021qlt}
R.~Abbott \textit{et al.} (LIGO Scientific, KAGRA and Virgo Collaborations),
Observation of Gravitational Waves from Two Neutron Star\textendash{}Black Hole Coalescences,
Astrophys. J. Lett. \textbf{915}, L5 (2021).
%doi:10.3847/2041-8213/ac082e
%[arXiv:2106.15163 [astro-ph.HE]].

\bibitem{gw150914-testGR}
B.~P.~Abbott \textit{et al.} (LIGO Scientific and Virgo Collaborations),,
Tests of general relativity with GW150914,
Phys. Rev. Lett. \textbf{116}, 221101 (2016).
%[arXiv:1602.03841 [gr-qc]].

\bibitem{gw170817-testGR}
B.~P.~Abbott \textit{et al.}  (LIGO Scientific and Virgo Collaborations), 
Tests of General Relativity with GW170817,
Phys. Rev. Lett. \textbf{123}, 011102 (2019).
% doi:10.1103/PhysRevLett.123.011102
%[arXiv:1811.00364 [gr-qc]].

\bibitem{gw170817-speed}
B.~P.~Abbott \textit{et al.} [LIGO Scientific, Virgo, Fermi-GBM and INTEGRAL],
Gravitational Waves and Gamma-rays from a Binary Neutron Star Merger: GW170817 and GRB 170817A,
Astrophys. J. Lett. \textbf{848}, L13 (2017).
%[arXiv:1710.05834 [astro-ph.HE]].

\bibitem{testGR_GWTC1}
B.~P.~Abbott \textit{et al.} [LIGO Scientific and Virgo Collaborations],
Tests of General Relativity with the Binary Black Hole Signals from the LIGO-Virgo Catalog GWTC-1,
Phys. Rev. D \textbf{100}, 104036 (2019).
%doi:10.1103/PhysRevD.100.104036
%[arXiv:1903.04467 [gr-qc]].


\bibitem{testGR_GWTC2}
R.~Abbott \textit{et al.}  (LIGO Scientific and Virgo Collaborations),
Tests of general relativity with binary black holes from the second LIGO-Virgo gravitational-wave transient catalog,
Phys. Rev. D \textbf{103}, 122002 (2021).
%doi:10.1103/PhysRevD.103.122002
%[arXiv:2010.14529 [gr-qc]].


\bibitem{MG1}
G.~Cognola, E.~Elizalde, S.~Nojiri, S.~D.~Odintsov, and S.~Zerbini,
Dark energy in modified Gauss-Bonnet gravity: Late-time acceleration and the hierarchy problem,
Phys. Rev. D \textbf{73}, 084007 (2006).
%doi:10.1103/PhysRevD.73.084007
%[arXiv:hep-th/0601008 [hep-th]].


\bibitem{MG2}
E.~J.~Copeland, M.~Sami, and S.~Tsujikawa,
Dynamics of dark energy,
Int. J. Mod. Phys. D \textbf{15}, 1753 (2006).
%doi:10.1142/S021827180600942X
%[arXiv:hep-th/0603057 [hep-th]].


\bibitem{MG3}
J.~Frieman, M.~Turner, and D.~Huterer,
Dark Energy and the Accelerating Universe,
Ann. Rev. Astron. Astrophys. \textbf{46}, 385 (2008).
%doi:10.1146/annurev.astro.46.060407.145243
%[arXiv:0803.0982 [astro-ph]].


\bibitem{MG4}
M.~Li, X.~D.~Li, S.~Wang, and Y.~Wang,
Dark Energy,
Commun. Theor. Phys. \textbf{56}, 525 (2011).
%doi:10.1088/0253-6102/56/3/24
%[arXiv:1103.5870 [astro-ph.CO]].


\bibitem{parity_violation}
T.~D. Lee and C.~N. Yang, %
Question of Parity Conservation in Weak Interactions,
Phys. Rev. \textbf{104}, 254 (1956).


\bibitem{cs1}
K.~Fujikawa,
Path Integral Measure for Gauge Invariant Fermion Theories,
Phys. Rev. Lett. \textbf{42}, 1195 (1979).
%doi:10.1103/PhysRevLett.42.1195

\bibitem{cs2}
K.~Fujikawa,
Path Integral for Gauge Theories with Fermions,
Phys. Rev. D \textbf{21}, 2848 (1980)
[erratum: Phys. Rev. D \textbf{22}, 1499 (1980)].
%doi:10.1103/PhysRevD.21.2848


\bibitem{cs_review}
S.~Alexander and N.~Yunes,
Chern-Simons Modified General Relativity,
Phys. Rept. \textbf{480}, 1 (2009).
%doi:10.1016/j.physrep.2009.07.002
%[arXiv:0907.2562 [hep-th]].

\bibitem{chern-simons1}
R.~Jackiw and S.~Y.~Pi,
Chern-Simons modification of general relativity,
Phys. Rev. D \textbf{68}, 104012 (2003).
%doi:10.1103/PhysRevD.68.104012
%[arXiv:gr-qc/0308071 [gr-qc]].

\bibitem{chern-simons2}
%\bibitem{Yunes:2010yf}
N.~Yunes, R.~O'Shaughnessy, B.~J.~Owen, and S.~Alexander,
Testing gravitational parity violation with coincident gravitational waves and short gamma-ray bursts,
Phys. Rev. D \textbf{82}, 064017 (2010).
%doi:10.1103/PhysRevD.82.064017
%[arXiv:1005.3310 [gr-qc]].


\bibitem{chern-simons3}
K.~Yagi, N.~Yunes, and T.~Tanaka,
Gravitational Waves from Quasi-Circular Black Hole Binaries in Dynamical Chern-Simons Gravity,
Phys. Rev. Lett. \textbf{109}, 251105 (2012)
[erratum: Phys. Rev. Lett. \textbf{116}, no.16, 169902 (2016); erratum: Phys. Rev. Lett. \textbf{124}, no.2, 029901 (2020)].
% doi:10.1103/PhysRevLett.116.169902
%[arXiv:1208.5102 [gr-qc]].


\bibitem{chern-simons4}
S.~H.~Alexander and N.~Yunes,
Gravitational wave probes of parity violation in compact binary coalescences,
Phys. Rev. D \textbf{97}, 064033 (2018).
%doi:10.1103/PhysRevD.97.064033
%[arXiv:1712.01853 [gr-qc]].


\bibitem{chern-simons5}
K.~Yagi and H.~Yang,
Probing Gravitational Parity Violation with Gravitational Waves from Stellar-mass Black Hole Binaries,
Phys. Rev. D \textbf{97}, 104018 (2018).
%doi:10.1103/PhysRevD.97.104018
%[arXiv:1712.00682 [gr-qc]].


\bibitem{Conroy}
A.~Conroy and T.~Koivisto,
Parity-Violating Gravity and GW170817 in Non-Riemannian Cosmology,
JCAP \textbf{12}, 016 (2019).
%doi:10.1088/1475-7516/2019/12/016
%[arXiv:1908.04313 [gr-qc]].


\bibitem{horava1}
P.~Horava,
Quantum Gravity at a Lifshitz Point,
Phys. Rev. D \textbf{79}, 084008 (2009).
%doi:10.1103/PhysRevD.79.084008
%[arXiv:0901.3775 [hep-th]].

\bibitem{horava2}
T.~Takahashi and J.~Soda,
Chiral Primordial Gravitational Waves from a Lifshitz Point,
Phys. Rev. Lett. \textbf{102}, 231301 (2009).
%doi:10.1103/PhysRevLett.102.231301
%[arXiv:0904.0554 [hep-th]].

\bibitem{horava3}
D.~Yoshida and J.~Soda,
Exploring the string axiverse and parity violation in gravity with gravitational waves,
Int. J. Mod. Phys. D \textbf{27}, 1850096 (2018); 
%doi:10.1142/S0218271818500967
%[arXiv:1708.09592 [gr-qc]]; 
M.~Satoh and J.~Soda,
Higher Curvature Corrections to Primordial Fluctuations in Slow-roll Inflation,
JCAP \textbf{09}, 019 (2008);
%doi:10.1088/1475-7516/2008/09/019
%[arXiv:0806.4594 [astro-ph]]; 
M.~Satoh, S.~Kanno and J.~Soda,
Circular Polarization of Primordial Gravitational Waves in String-inspired Inflationary Cosmology,
Phys. Rev. D \textbf{77}, 023526 (2008);
%doi:10.1103/PhysRevD.77.023526
%[arXiv:0706.3585 [astro-ph]]; 
J.~Soda, H.~Kodama, and M.~Nozawa,
Parity Violation in Graviton Non-gaussianity,
JHEP \textbf{08}, 067 (2011);
%doi:10.1007/JHEP08(2011)067
%[arXiv:1106.3228 [hep-th]]; 
T.~Zhu, Q.~Wu, A.~Wang, and F.~W.~Shu,
U(1) symmetry and elimination of spin-0 gravitons in Horava-Lifshitz gravity without the projectability condition,
Phys. Rev. D \textbf{84}, 101502 (2011).
%doi:10.1103/PhysRevD.84.101502
%[arXiv:1108.1237 [hep-th]].

\bibitem{horava4}
A.~Wang, Q.~Wu, W.~Zhao, and T.~Zhu,
Polarizing primordial gravitational waves by parity violation,
Phys. Rev. D \textbf{87}, 103512 (2013).
%doi:10.1103/PhysRevD.87.103512
%[arXiv:1208.5490 [astro-ph.CO]].


\bibitem{horava5}
T.~Zhu, W.~Zhao, Y.~Huang, A.~Wang, and Q.~Wu,
Effects of parity violation on non-gaussianity of primordial gravitational waves in Ho\v{r}ava-Lifshitz gravity,
Phys. Rev. D \textbf{88}, 063508 (2013).
%doi:10.1103/PhysRevD.88.063508


\bibitem{horava6}
A.~Wang,
Ho\v{r}ava gravity at a Lifshitz point: A progress report,
Int. J. Mod. Phys. D \textbf{26}, 1730014 (2017).
%doi:10.1142/S0218271817300142
%[arXiv:1701.06087 [gr-qc]].


\bibitem{chiral_ST}
M.~Crisostomi, K.~Noui, C.~Charmousis, and D.~Langlois,
Beyond Lovelock gravity: Higher derivative metric theories,
Phys. Rev. D \textbf{97}, 044034 (2018).
%doi:10.1103/PhysRevD.97.044034
%[arXiv:1710.04531 [hep-th]].


\bibitem{chiral_ST1}
A.~Nishizawa and T.~Kobayashi, Parity-violating gravity and GW170817,
Phys. Rev. D \textbf{98}, 124018 (2018).
%doi:10.1103/PhysRevD.98.124018
%[arXiv:1809.00815 [gr-qc]].


\bibitem{chiral_ST2}
X.~Gao and X.~Y.~Hong,
Propagation of gravitational waves in a cosmological background,
Phys. Rev. D \textbf{101}, 064057 (2020).
%doi:10.1103/PhysRevD.101.064057
%[arXiv:1906.07131 [gr-qc]].


\bibitem{SME1}
V.~A.~Kosteleck\'y and M.~Mewes,
Testing local Lorentz invariance with gravitational waves,
Phys. Lett. B \textbf{757}, 510 (2016).
%doi:10.1016/j.physletb.2016.04.040
%[arXiv:1602.04782 [gr-qc]].


\bibitem{SME2}
Q.~G.~Bailey and V.~A.~Kostelecky,
Signals for Lorentz violation in post-Newtonian gravity,
Phys. Rev. D \textbf{74}, 045001 (2006).
%doi:10.1103/PhysRevD.74.045001
%[arXiv:gr-qc/0603030 [gr-qc]].


\bibitem{SME3}
M.~Mewes,
Signals for Lorentz violation in gravitational waves,
Phys. Rev. D \textbf{99}, 104062 (2019).
%doi:10.1103/PhysRevD.99.104062
%[arXiv:1905.00409 [gr-qc]].



\bibitem{SME4}
L.~Shao,
Combined search for anisotropic birefringence in the gravitational-wave transient catalog GWTC-1,
Phys. Rev. D \textbf{101}, 104019 (2020).
%doi:10.1103/PhysRevD.101.104019
%[arXiv:2002.01185 [hep-ph]].


\bibitem{SME5}
Z.~Wang, L.~Shao and C.~Liu,
New limits on the Lorentz/CPT symmetry through fifty gravitational-wave events,
Astrophysics. J. {\bf 921}, 158 (2021).
%doi:10.3847/1538-4357/ac223c
%[arXiv:2108.02974 [gr-qc]].


\bibitem{PGW1}
A.~Lue, L.~M.~Wang and M.~Kamionkowski,
Cosmological signature of new parity violating interactions,
Phys. Rev. Lett. \textbf{83}, 1506 (1999).
%doi:10.1103/PhysRevLett.83.1506
%[arXiv:astro-ph/9812088 [astro-ph]].

\bibitem{PGW2}
S.~Alexander and J.~Martin,
Birefringent gravitational waves and the consistency check of inflation,
Phys. Rev. D \textbf{71}, 063526 (2005).
%doi:10.1103/PhysRevD.71.063526
%[arXiv:hep-th/0410230 [hep-th]].



\bibitem{PGW3}
N.~Bartolo, L.~Caloni, G.~Orlando and A.~Ricciardone,
Tensor non-Gaussianity in chiral scalar-tensor theories of gravity,
JCAP \textbf{03}, 073 (2021).
%doi:10.1088/1475-7516/2021/03/073
%[arXiv:2008.01715 [astro-ph.CO]].

\bibitem{Fu:2020tlw}
C.~Fu, J.~Liu, T.~Zhu, H.~Yu and P.~Wu,
Resonance instability of primordial gravitational waves during inflation in Chern\textendash{}Simons gravity,
Eur. Phys. J. C \textbf{81}, 204 (2021).
%doi:10.1140/epjc/s10052-021-09001-2
%[arXiv:2006.03771 [gr-qc]].



\bibitem{Okounkova:2021xjv}
M.~Okounkova, W.~M.~Farr, M.~Isi and L.~C.~Stein,
Constraining gravitational wave amplitude birefringence and Chern-Simons gravity with GWTC-2,
arXiv:2101.11153 [gr-qc].


\bibitem{Hu:2020rub}
Q.~Hu, M.~Li, R.~Niu and W.~Zhao,
Joint Observations of Space-based Gravitational-wave Detectors: Source Localization and Implication for Parity-violating gravity,
Phys. Rev. D \textbf{103}, 064057 (2021).
%doi:10.1103/PhysRevD.103.064057
%[arXiv:2006.05670 [gr-qc]].

\bibitem{CS_gb}
R.~Nair, S.~Perkins, H.~O.~Silva and N.~Yunes,
Fundamental Physics Implications for Higher-Curvature Theories from Binary Black Hole Signals in the LIGO-Virgo Catalog GWTC-1,
Phys. Rev. Lett. \textbf{123}, 191101 (2019).
%doi:10.1103/PhysRevLett.123.191101
%[arXiv:1905.00870 [gr-qc]].



\bibitem{sai_wang}
S.~Wang and Z.~C.~Zhao,
Tests of CPT invariance in gravitational waves with LIGO-Virgo catalog GWTC-1,
Eur. Phys. J. C \textbf{80}, 1032 (2020).
%doi:10.1140/epjc/s10052-020-08628-x
%[arXiv:2002.00396 [gr-qc]].

\bibitem{tanaka}
K.~Yamada and T.~Tanaka,
Parametrized test of parity-violating gravity using GWTC-1 events,
PTEP \textbf{2020}, 093E01 (2020) 
%doi:10.1093/ptep/ptaa103
[arXiv:2006.11086 [gr-qc]].


\bibitem{yi-fan1}
Y.~F.~Wang, R.~Niu, T.~Zhu, and W.~Zhao,
Gravitational Wave Implications for the Parity Symmetry of Gravity in the High Energy Region,
Astrophys. J. \textbf{908}, 58 (2021)
%doi:10.3847/1538-4357/abd7a6
[arXiv:2002.05668 [gr-qc]].


\bibitem{Wang:2021gqm}
Y.~F.~Wang, S.~M.~Brown, L.~Shao, and W.~Zhao,
Tests of Gravitational-Wave Birefringence with the Gravitational-Wave Catalog,
arXiv:2109.09718 [astro-ph.HE].

\bibitem{waveform}
W. Zhao, T. Zhu, J. Qiao, and A. Wang,  Waveform of gravitational waves in the general parityviolating gravities, Phys. Rev. D \textbf{101}, 024002 (2020);  
J.~Qiao, T.~Zhu, W.~Zhao, and A.~Wang, Waveform of gravitational waves in the ghost-free parity-violating gravities, Phys. Rev. D \textbf{100}, 124058 (2019) [arXiv:1909.03815 [gr-qc]].


\bibitem{Li:2020xjt}
M.~Li, H.~Rao, and D.~Zhao,
A simple parity violating gravity model without ghost instability,
JCAP \textbf{11}, 023 (2020)
%doi:10.1088/1475-7516/2020/11/023
[arXiv:2007.08038 [gr-qc]].


\bibitem{Li:2021wij}
M.~Li, H.~Rao, and Y.~Tong,
Revisiting a parity violating gravity model without ghost instability: Local Lorentz covariance,
Phys. Rev. D \textbf{104}, 084077 (2021)
[arXiv:2104.05917 [gr-qc]].

\bibitem{TEGR}
J.G. Pereira and R. Aldrovandi, Teleparallel Gravity: An Introduction (Springer, Dordrecht, 2013).

\bibitem{Bahamonde:2021gfp}
S.~Bahamonde, K.~F.~Dialektopoulos, C.~Escamilla-Rivera, G.~Farrugia, V.~Gakis, M.~Hendry, M.~Hohmann, J.~L.~Said, J.~Mifsud, and E.~Di Valentino, Teleparallel Gravity: From Theory to Cosmology,
[arXiv:2106.13793 [gr-qc]].

\bibitem{NY-1}
S.~Mercuri,
Peccei-Quinn mechanism in gravity and the nature of the Barbero-Immirzi parameter,
Phys. Rev. Lett. \textbf{103}, 081302 (2009).
%doi:10.1103/PhysRevLett.103.081302
%[arXiv:0902.2764 [gr-qc]].

\bibitem{NY-2}
O.~Castillo-Felisola, C.~Corral, S.~Kovalenko, I.~Schmidt and V.~E.~Lyubovitskij,
Axions in gravity with torsion,
Phys. Rev. D \textbf{91}, 085017 (2015).
%doi:10.1103/PhysRevD.91.085017
%[arXiv:1502.03694 [hep-ph]].

\bibitem{NY-3}
R.~D.~Peccei and H.~R.~Quinn,
CP Conservation in the Presence of Instantons,
Phys. Rev. Lett. \textbf{38}, 1440 (1977).

\bibitem{PPN}
H.~Rao,
Parameterized post-Newtonian limit of the Nieh-Yan modified teleparallel gravity,
arXiv:2107.08597 [gr-qc].



\bibitem{Bombacigno:2021bpk}
F.~Bombacigno, S.~Boudet, G.~J.~Olmo, and G.~Montani,
Big bounce and future time singularity resolution in Bianchi I cosmologies: The projective invariant Nieh-Yan case,
Phys. Rev. D \textbf{103}, 124031 (2021).
% doi:10.1103/PhysRevD.103.124031
%[arXiv:2105.06870 [gr-qc]].

\bibitem{Bahamonde:2017wwk}
S.~Bahamonde, C.~G.~B\"ohmer and M.~Kr\v{s}\v{s}\'ak,
New classes of modified teleparallel gravity models,
Phys. Lett. B \textbf{775}, 37 (2017).
%doi:10.1016/j.physletb.2017.10.026
%[arXiv:1706.04920 [gr-qc]].

\bibitem{Bahamonde:2015zma}
S.~Bahamonde, C.~G.~B\"ohmer, and M.~Wright,
Modified teleparallel theories of gravity,
Phys. Rev. D \textbf{92}, 104042 (2015).
%doi:10.1103/PhysRevD.92.104042
%[arXiv:1508.05120 [gr-qc]].

\bibitem{Hohmann:2020dgy}
M.~Hohmann and C.~Pfeifer,
Teleparallel axions and cosmology,
Eur. Phys. J. C \textbf{81}, 376 (2021).
%doi:10.1140/epjc/s10052-021-09165-x
%[arXiv:2012.14423 [gr-qc]].

%\cite{Zhang:2021kqn}
\bibitem{Zhang:2021kqn}
Y.~Zhang and H.~Zhang,
Distinguish the $f(T)$ model from $\Lambda $CDM model with Gravitational Wave observations,
Eur. Phys. J. C \textbf{81}, 706 (2021).
%doi:10.1140/epjc/s10052-021-09501-1
%[arXiv: 2108.05736 [astro-ph.CO]].




\bibitem{Li:2021mdp}
M.~Li and D.~Zhao,
A simple parity violating model in the symmetric teleparallel gravity and its cosmological perturbations,
arXiv:2108.01337 [gr-qc].

\bibitem{SMExx}
V.~A.~Kosteleck\'y and M.~Mewes,
Lorentz and Diffeomorphism Violations in Linearized Gravity,
Phys. Lett. B \textbf{779}, 136 (2018).
%doi:10.1016/j.physletb.2018.01.082
%[arXiv:1712.10268 [gr-qc]].

\bibitem{SME-EM}
V.~A.~Kostelecky and M.~Mewes,
Electrodynamics with Lorentz-violating operators of arbitrary dimension,
Phys. Rev. D \textbf{80}, 015020 (2009).
%doi:10.1103/PhysRevD.80.015020
%[arXiv:0905.0031 [hep-ph]].

\bibitem{PGG_pv}
R. Kuhfuss and J. Nitsch, Propagating Modes in Gauge Field Theories of Gravity, Gen. Relat. Gravit. {\bf 18}, 1207 (1986).


\bibitem{Jimenez:2021nup}
J.~B.~Jim\'enez and T.~S.~Koivisto,
Noether charges in the geometrical trinity of gravity,
arXiv:2111.04716 [gr-qc].

\bibitem{strong}
A. Jim\"enez Cano, {\em METRIC-AFFINE GAUGE THEORIES OF GRAVITY: Foundations and new insights},” Granada U. (2021).

\bibitem{Ong:2013qja}
Y.~C.~Ong, K.~Izumi, J.~M.~Nester, and P.~Chen,
Problems with Propagation and Time Evolution in f(T) Gravity,
Phys. Rev. D \textbf{88}, 024019 (2013).
%doi:10.1103/PhysRevD.88.024019
%[arXiv:1303.0993 [gr-qc]].

\bibitem{DeFelice:2012mx}
A.~De Felice, A.~E.~Gumrukcuoglu, and S.~Mukohyama,
Massive gravity: nonlinear instability of the homogeneous and isotropic universe,
Phys. Rev. Lett. \textbf{109}, 171101 (2012).
%doi:10.1103/PhysRevLett.109.171101
%[arXiv:1206.2080 [hep-th]].

\bibitem{Jimenez:2020gbw}
J.~B.~Jim\'enez and A.~Jim\'enez-Cano,
On the strong coupling of Einsteinian Cubic Gravity and its generalisations,
JCAP \textbf{01}, 069 (2021)
%doi:10.1088/1475-7516/2021/01/069
[arXiv:2009.08197 [gr-qc]].

\bibitem{GR_wave}
P.~Jaranowski, A.~Krolak, and B.~F.~Schutz,
Data analysis of gravitational - wave signals from spinning neutron stars. 1. The Signal and its detection,
Phys. Rev. D \textbf{58}, 063001 (1998); 
W.~Zhao and L.~Wen,
Localization accuracy of compact binary coalescences detected by the third-generation gravitational-wave detectors and implication for cosmology,
Phys. Rev. D \textbf{97}, 064031 (2018).


\bibitem{GR_wave2}
B.~S.~Sathyaprakash and B.~F.~Schutz,
Physics, Astrophysics and Cosmology with Gravitational Waves,
Living Rev. Rel. \textbf{12}, 2 (2009).

\bibitem{Schmidt:2014iyl}
P.~Schmidt, F.~Ohme, and M.~Hannam,
Towards models of gravitational waveforms from generic binaries II: Modelling precession effects with a single effective precession parameter,
Phys. Rev. D \textbf{91}, 024043 (2015)
[arXiv:1408.1810 [gr-qc]].

\bibitem{Hannam:2013oca}
M.~Hannam, P.~Schmidt, A.~Boh\'e, L.~Haegel, S.~Husa, F.~Ohme, G.~Pratten, and M.~P\"urrer,
Simple Model of Complete Precessing Black-Hole-Binary Gravitational Waveforms,
Phys. Rev. Lett. \textbf{113}, 151101 (2014).

\bibitem{Pratten:2020ceb}
G.~Pratten, C.~Garc\'\i{}a-Quir\'os, M.~Colleoni, A.~Ramos-Buades, H.~Estell\'es, M.~Mateu-Lucena, R.~Jaume, M.~Haney, D.~Keitel, and J.~E.~Thompson, \textit{et al.}
Computationally efficient models for the dominant and subdominant harmonic modes of precessing binary black holes,
Phys. Rev. D \textbf{103}, 104056 (2021).


\bibitem{data_GW}
R.~Abbott \textit{et al.} (LIGO Scientific and Virgo Collaborations),
Open data from the first and second observing runs of Advanced LIGO and Advanced Virgo,
SoftwareX \textbf{13}, 100658 (2021)
[arXiv:1912.11716 [gr-qc]].


\bibitem{data_GW1}
B. P. Abbott \textit{et al.} (LIGO Scientific and Virgo Collaborations), A Guide to LIGO–Virgo Detector Noise and Extraction of Transient Gravitational-Wave Signals, Classical Quantum Gravity {\bf 37}, 055002 (2020). 

\bibitem{data_GW2}
B.~P.~Abbott \textit{et al.} (LIGO Scientific and Virgo Collaborations),
GWTC-2 Data Release: Parameter Estimation Samples and Skymaps," 
https://dcc.ligo.org/LIGO- P2000223/public.


\bibitem{PSD1}
N.~J.~Cornish and T.~B.~Littenberg,
BayesWave: Bayesian Inference for Gravitational Wave Bursts and Instrument Glitches,
Class. Quant. Grav. \textbf{32}, 135012 (2015).


\bibitem{PSD2}
T.~B.~Littenberg and N.~J.~Cornish,
Bayesian inference for spectral estimation of gravitational wave detector noise,
Phys. Rev. D \textbf{91}, 084034 (2015).


\bibitem{bilby}
G.~Ashton, M.~H\"ubner, P.~D.~Lasky, C.~Talbot, K.~Ackley, S.~Biscoveanu, Q.~Chu, A.~Divakarla, P.~J.~Easter, and B.~Goncharov, \textit{et al.}
BILBY: A user-friendly Bayesian inference library for gravitational-wave astronomy,
Astrophys. J. Suppl. \textbf{241}, 27 (2019)
[arXiv:1811.02042 [astro-ph.IM]]; 
I.~M.~Romero-Shaw, C.~Talbot, S.~Biscoveanu, V.~D'Emilio, G.~Ashton, C.~P.~L.~Berry, S.~Coughlin, S.~Galaudage, C.~Hoy, and M.~H\"ubner, \textit{et al.}
Bayesian inference for compact binary coalescences with bilby: validation and application to the first LIGO\textendash{}Virgo gravitational-wave transient catalogue,
Mon. Not. Roy. Astron. Soc. \textbf{499}, 3295-3319 (2020)
[arXiv:2006.00714 [astro-ph.IM]].



\end{thebibliography}
\end{document}